\documentclass[lettersize,journal]{IEEEtran}
\usepackage{url}
\usepackage{graphicx}
\usepackage{cite}
\usepackage{hyperref}
\usepackage{xcolor}
\usepackage{textcomp}
\usepackage{stfloats}
\usepackage{url}
\usepackage{verbatim}
\usepackage{array,multirow}
\usepackage{makecell}
\usepackage[caption=false,font=normalsize,labelfont=sf,textfont=sf]{subfig}
\begin{document}

\title{Security Attacks and Solutions for Digital Twins}

\author{Sabah Suhail, Raja Jurdak,~\IEEEmembership{Senior~Member,~IEEE}, and Rasheed Hussain,~\IEEEmembership{Senior~Member,~IEEE}
\thanks{S. Suhail is with Vienna University of Economics and Business, Austria (e-mail: sabah.suhail@wu.ac.at.}
\thanks{R. Jurdak is with Queensland University of Technology, Australia (e-mail: r.jurdak@qut.edu.au).}
\thanks{R. Hussain is with the Bristol Digital Futures Institute (BDFI) and Smart Internet Lab, University of Bristol, UK (e-mail: rasheed.hussain@bristol.ac.uk).}
}
\maketitle

\begin{abstract}
Digital twins, being the virtual replicas of their physical counterparts, share valuable knowledge of the underlying system. Therefore, they might become a potential source of data breaches and a playground for attackers to launch covert attacks.
It is imperative to investigate necessary countermeasures to mitigate such attacks.

\end{abstract}


\IEEEpeerreviewmaketitle

\section{Introduction}
\IEEEPARstart{T}{}he convergence of Information Technology (IT) and Operational Technology (OT) in Industrial Control Systems (ICSs) will speed up the realization of both Industry 4.0 and 5.0 
~\cite{dietz2020unleashing}. At the core of Industry 4.0 and beyond, are the Cyber-Physical Systems (CPSs) that connect physical (OT) and cyber (IT) components through computational and networking capabilities~\cite{Eckhart2019}. ICSs are a subset of CPSs. ICSs provide promising solutions to various industrial ecosystems but they substantially expand the attack surface~\cite{suhail2021blockchainbased}. By exploiting different attack vectors (i.e., cyber and physical), attackers can launch Advanced Persistent Threats (APTs) through which they covertly reside in the system to continually exfiltrate information or undermine the critical processes. To ensure that the system operates securely and safely, we need essential measures to secure CPSs: (1) evaluating the operational behavior of the system, and (2) conducting penetration testing on the system to identify vulnerabilities or threats~\cite{suhail2021securing}. As CPSs cannot be deactivated for carrying out such analysis, assessing the system's security level requires online solutions that accurately reflect the actual CPS operations while avoiding any interference or side-effects of testing on the live systems~\cite{suhail2021securing}. Digital Twins are one such promising solution~\cite{suhail2021blockchainbased} which addresses this constraint.

Digital twin is a virtual replica of a physical asset (device or process) that enables analyzing, predicting, and optimizing operations by utilizing real-time and historical data~\cite{suhail2021securing}. In the information security domain, digital twins strengthen the security of CPS through various security-enhancing use cases, including system training and testing, security testing, and detecting system misconfigurations~\cite{eckhart2018towards}. To do so, digital twins run synchronously with their physical counterparts where the goal is to track~\emph{data inconsistencies} between the physical and the virtual entity~\cite{suhail2020trustworthy}. Digital twins collect and integrate data from multiple sources, such as sensory data from the physical environment, historical data from CPS lifecycle phases, and domain knowledge from experts, to learn the emergent behavior of the physical environment, which serves as valuable insights for anomaly detection. Then, following a closed-loop, we feed the optimized data to the physical entity to adapt operations to latest state operations. 

Despite various digital twins use cases that could reinforce security in system engineering or during the operation phase of CPS~\cite{Eckhart2019}, the emergence of stealthy threats allows attackers to exploit digital twins to launch attacks on the CPS. Digital twins, being the virtual (digital) replicas of their physical counterparts, share functional requirements and operational behavior of the underlying systems. Therefore, digital twins may act as a potential source of data breaches, leading to the abuse case of digital twins~\cite{Eckhart2019}. Attackers may exploit the deep 
knowledge about the physical process and devices accessible through digital twins with a two-stage strategy: use the key input data source namely, digital twins into a malicious state, and then through that state manipulate the underlying physical system’s behavior covertly~\cite{suhail2021securing}. For example, manipulating the behavior of digital twins by modifying their defined states which would correspond to a direct attack on field devices, particularly when automated feedback loops are enabled between the physical objects and their digital counterparts~\cite{Eckhart2019}. It is necessary to ensure the trustworthiness of digital twins for timely corrections because ignoring such pre-emptive measures may lead to a feedback loop of erroneous data into the system resulting in the Garbage In Garbage Out (GIGO) problem~\cite{suhail2020trustworthy}. Moreover, in human-machine collaboration scenarios, a slight system dysfunction caused by mirroring of malicious replicas may pose a severe threat to human safety. The repercussions of exploiting digital twins may have severe consequences within the {\it digital thread} that links data across multiple digital twin instances or CPS lifecycle phases.
These links are an attractive target for attacks because the entire product lifecycle can be targeted in a data breach~\cite{suhail2021blockchainbased} such as when manipulating high-valued design artifacts. Furthermore, attack on digital thread may affect the next generation CPS system where digital twins data can be used as historical data. The authors of~\cite{Eckhart2019} have shown how digital twins can be exploited for launching attacks as one of the open research directions. However, to the best of our knowledge, there are no previous studies on the abuse case of digital twins. Our main research contributions are as follows:
\begin{itemize}
\item We discuss possible attacks on digital twins where the attacker defeats the security of digital twins by either manipulating the benign behavior of digital twins or by exploiting the cyclic state update from the physical process to digital twins to steer the CPS into an insecure state. 
\item We discuss potential security solutions that can mitigate the possible attacks (we have identified) on digital twins. 
\item In this context, we propose a digital twin-based gamification approach that can access the security level of the digital twins. Furthermore, the gamification approach provides security analysts with a controlled, supportive virtual training environment.
\end{itemize}

\section{Digital Twins Abuse Case : An Attacker's Perspective}
To understand the anatomy of a cyberattack, we need to understand the adversary tactics (as Table~\ref{tab:attack_ideas} shows). The following section discusses different attackers' strategies on digital twins.


\begin{table*}[!t]
  \centering
\caption{Attack on digital twins: an attacker's perspective.} \label{tab:attack_ideas}
 \renewcommand*{\arraystretch}{1.3}
  \begin{tabular}{|p{2.70cm}|p{11.50cm}|} 
  \hline
\thead{\textbf{Attacker's artifacts}} & \thead{\textbf{Goals}}
 \\
\hline
\multirow{3}{*}{Product lifecycle} & 
 \begin{itemize}
    \item Manipulate benign behavior of digital twins to steer the CPS into an insecure state
    \item Exploit digital thread as it links data throughout the entire product lifecycle
\end{itemize} 
\\
Replication mode & \hspace*{4.5
 mm} \hangindent=2.0em Run direct cyclic state updates by replicating the
virtual behavior of digital twins to the corresponding program states of physical devices 
\\
\multirow{3}{*}{Simulation mode} &
 \begin{itemize}
     \item Learn system behavior by re-running test simulations
      \item Manipulate simulation parameters or system specifications’ data during security tests
 \end{itemize}   
 \\
 Design phase & \hspace*{4.5
 mm} \hangindent=2.0em Exploit specification-based or machine learning-based process knowledge of digital twins 
 \\
\multirow{3}{*}{Decommissioning phase} & \begin{itemize}
    \item Reuse system’s knowledge due to improper disposal of the digital twins
    \item Use data security breach such as unauthorized access to gain access to archived digital twins’ data
\end{itemize} 
\\
\multirow{3}{*}{Lateral movement} & 
\begin{itemize}
    \item Gain control over high-value assets such as design artifacts
    \item Manipulate sensor readings or simulation parameters at random intervals
    while ensuring that the new values do not deviate significantly from the real process values
\end{itemize} 
 \\
 \hline
\end{tabular}
\end{table*}

\subsection{Reconnaissance attacks}\label{rec}
Reconnaissance involves intelligence gathering. This is achieved through activities such as network scanning, exploiting zero-day vulnerabilities, and enumerating services to identify security loopholes in the underlying system. For instance, the Triton malware targeted a petrochemical plant in Saudi Arabia and gained a foothold in the IT/OT networks to target Safety Instrumented Systems (SIS). However, the attacker may go beyond conventional network reconnaissance in industrial ecosystems to achieve the desired objectives. Sophisticated malware can defeat isolation mechanisms, including air gaps, sandboxes, virtualization, and so on. For example, the Stuxnet malware aimed at a Uranium enrichment plant demonstrates how to overcome air gaps~\cite{langner2011stuxnet}. Thus, beginning with reconnaissance scans, the attacker may gather information about the loopholes in the infrastructure and then use Stuxnet- or Triton-inspired malware strategies to launch attacks on digital twins. 

\subsection{Which digital twins mode could cause more damage to the system?}
\label{sim-rep}
Digital twins do not need to replicate the CPS in its entirety~\cite{Eckhart2019}. 
Given that the virtual representation in digital twins mimics the functionality of corresponding processes or devices with enough details reasonable feature generalizations or simplifications can occur if they stay context-aware~\cite{minerva2021digital}. More precisely, the goal of building digital twins is to provide a cost-effective solution to test the physical system rather than replicating (in terms of simulation or emulation) the system. Nevertheless, accurate representation of digital twins also contributes to the likelihood of a successful attack. In this case, the attacker considers the operation modes of digital twins. Next, we discuss the operation modes of digital twins which could be exploited by attackers. In replication mode, we record the event in a real system and then replay it while emulating the system behavior~\cite{dietz2020unleashing}. To do so, twins and their physical counterparts must be synchronized through sensor measurements, network communication, or log files~\cite{suhail2021securing}. We maintain a constant connection with the physical counterpart by integrating the system speciﬁcation and the current state’s data. The replication mode may initiate direct cyclic updates to and from the digital twins. However, to use the replication mode, the attacker may follow implicit or explicit attacking points and needs to stay active to avoid the problem of time-dependent state synchronization and consistency of information between the physical entity and its different replicas.

Simulation mode operates in an isolated virtual environment without having a direct connection to the live systems. It requires user-specified settings and parameters as input. In simulation mode, being reproducible and repeatable with a broad range of trial-and-error learning mechanisms~\cite{dietz2020unleashing}, can be directly used or tailored according to the attacker's needs. The attacker can reveal emergent system behaviors by resetting and re-running the simulation until he/she achieves his/her insidious goals. Even worse, it can target the theme of simulation mode - security by design by reversing the defined configurations during security tests within the virtual environment. Furthermore, the attacker can learn the system state passively. However, since the simulation mode runs independently of its physical counterpart, the attacker cannot trigger automated attacks on the system due to the absence of a direct feedback loop.

\subsection{Victimizing physical system or twin?} \label{physical-virtual}
\subsubsection{Targeting the physical system}
Integrating general-purpose IT systems with ICSs introduces novel attack vectors~\cite{dietz2020unleashing}. Usually operational functionality outweighs security, therefore loopholes in the system infrastructure allow attackers to launch advanced covert attacks (e.g., APTs). ICS-tailored malware (for instance, Stuxnet), is one such example wherein a malicious code self-protects and self-updates itself while intercepting and modifying the data sent to and from Programmable Logic Controllers (PLCs) to compromise its target covertly~\cite{langner2011stuxnet}. Despite the high level of effort needed, attackers may choose to directly interact with the physical system (through industrial espionage or a cyberwarfare weapon) with the objective to damage or destroy it. 

\subsubsection{Targeting the digital twin}
When attackers choose digital twins as a target they can ultimately destroy the physical asset. This is because launching attacks on digital twins is like launching them on the physical system because malicious code can intercept and modify the simulation parameters in the digital twins of PLCs.

Next, we discuss motivations behind attacks on digital twins:
\begin{itemize}
      \item Digital twins can serve as an anomaly-based intrusion detection tool wherein through time and state, users (both benign and malicious) can continuously monitor the ongoing processes to observe the expected behavior of the system~\cite{suhail2021securing}. For instance, analyzing the relationships among dynamic variables (of the physical process) and historical variables (of the virtual process) facilitates the detection of Safety and Security (S\&S) rule violations. On the one hand, a benign user uses digital twins to spot deviations from a defined or learned baseline and alert security analysts~\cite{eckhart2018towards}. On the other hand, a malicious user can exploit the correlation of variables to disrupt the digital twins' behavior such 
    that the twins do not follow the defined 
    misbehavior (knowledge-based or behavior-based) pattern and thus unable to detect the misbehavior. As the attacker follows the ``living off the land" attack strategy (i.e., without using any of the illegitimate software and functions to perform malicious actions), no intrusion can be detected and thus, it is hard to spot long-term deviations.  
     \item Digital twins, from design to dismissal, are among the key input data sources for the physical systems. For example, the asset prototype is designed and tailored based on the simulation mode in the engineering phase. Then during the operation phase, the physical asset further evolves and optimizes its functionalities based on the simulation or replication modes of digital twins. With these chains of data inputs/outputs, exploiting digital twins may provide insights into attack vectors that can be used to plot long-term attacks on next generation CPS systems.
\end{itemize}

Digital twins must exhibit sufficient fidelity in terms of functionality and time-sensitive operational behavior of the physical component to protect against failures of the real system. Besides, to maintain backward compatibility, a corresponding evolution of twin is needed as the physical object evolves over time~\cite{rasheed2020digital}. With such challenges, attackers need to adapt their attack techniques to the new requirements of digital twins.

\subsubsection{Could the hybrid approach be worse than attacking either the physical or the digital entity?} \label{hybrid}
Considering the complexity and the ever-changing threat landscape, it is possible for the attacker covertly reside at both places, i.e., the physical system and its virtual counterpart. The proliferation of such sophisticated attacks enables cyber-attackers to conceal themselves within enterprise network traffic while actively hunting for valuable data. Even if they get caught while feeding adversarial data into the physical system, attackers may still exploit vulnerable entry points into digital twins.

\subsection{Morphing digital twins through lifecycle phases}\label{morphing}
\subsubsection{Engineering phase} \label{engineering}
The concept of building the process knowledge of digital twins can be achieved in two ways. First, by utilizing the CPS specification (such as the network and/or logic layer) to model the physical counterpart~\cite{eckhart2018towards}. Second, by utilizing machine learning to learn security-related aspects based on sensor data~\cite{groshev2021toward} without obtaining process knowledge from DTs~\cite{suhail2021securing}.
The process knowledge acquired through the specification of digital twins is less favorable compared to when it is based on machine learning because the former emulates the behavior of the system more closely. But complete transparency into the inner working of Artificial Intelligence (AI) models may expose them to adversarial attacks by allowing cyber attackers to make inferences from live cyber data or perform model poisoning into the training workflows.

\subsubsection{Decommissioning phase} \label{decomission}
Usually, due to obsolescence or replacement of outdated hardware/software, the asset and its digital twin are destroyed during the dismissal phase. Nevertheless, as the knowledge about the predecessor system can be rehashed, digital twin data could be backed up and procured by similar objects or domain experts to optimize the next generation of the system. No matter whether the digital twins are destroyed or retained for future usage, attackers might exploit them. For instance, digital twins can be attractive targets of data breach incidents if they are not carefully disposed while complying with proper media sanitization guidelines~\cite{Eckhart2019}. Similarly, if digital twins are archived for future usage without complying with adequate data security measures, they can be easy targets of security breaches.

\subsection{Lateral movement} \label{lateral}
By moving deeper into the system in search of sensitive information or gaining control over high-value assets, attackers strategically target specific sensors and manipulate readings~\cite{Eckhart2019}. 
More specifically, to make the attack more sophisticated, attackers may even consider process dynamics, i.e., the time-dependent behavior of a process in response to data input, to launch an attack.

\subsection{Drawbacks of desirable features of digital twins}\label{drawbacks}
During the operation phase of CPSs, cyclic state updates can be allowed from the digital twins to the physical process or vice versa. Although such actions aim to optimize the underlying operations, however, doing so, attacks initiated through digital twins may have similar repercussions as those attacks which are launched directly on real field devices. Similarly, ranging from low to high, the precise representation of digital twins, i.e., fidelity, is essential in the deployment of digital twins. 
Usually, the required level of fidelity, i.e., the degree of state or behavioral accuracy, for the digital twins is based on the underlying use cases or other fidelity metrics. For instance, virtual honeypots need to be more realistic to lure attackers, thus requiring high-fidelity~\cite{Eckhart2019}. However, cross-checking the engineering knowledge (such as 
verifying device data against threshold values, identifying unidentified connections or unknown devices, and so on~\cite{suhail2022towards}) based on the design specifications of the underlying CPS at the network and logic layer that includes physical devices~\cite{eckhart2018towards} can be achieved with low-fidelity digital twins. The design specifications of the underlying CPS at the network and logic layer~\cite{eckhart2018towards} may provide low-level fidelity, however, real-time sensor data provides high-level fidelity digital twins. Simply put, the more the digital twins accurately reflect their physical counterpart, the easier it is for the attacker to understand the system behavior.

\section{Countermeasures}
The following section discusses the countermeasures to thwart attacks discussed above on digital twins. 

\subsection{Blockchain-based digital twins}
Given that the digital twin data is used as an input source to the CPS physical processes, the digital twin must be built on trusted data~\cite{suhail2021securing}. In this context, empowering digital twins with blockchain allow industries to manage data on a distributed ledger while ensuring trusted digital twin data coordination across multiple stakeholders~\cite{suhail2021blockchainbased}.
Next, we discuss possible solutions that can mitigate the attacks we have identified earlier on digital twins.

\subsubsection{Orchestrating provenance}
In ICSs, the controllers usually focus on code syntactic disregard for changes originating from authorized engineering stations mistakenly or maliciously~\cite{langner2011stuxnet}. The inability to track changes in the system opens up opportunities for the wrongdoers to compromise targets covertly (such as reconnaissance attacks discussed in section~\ref{rec}) before launching an overt attack. Similarly, infrastructure vulnerabilities (such as missing or weak authentication and authorization credentials) lead to exploitation of process knowledge of digital twins (during the engineering phase~\ref{engineering}) and the decommissioning phase (mainly when digital twins are archived for future usage). Therefore, we need to enforce mechanisms that ensure trusted digital twins calls to keep track of activities such as granting access privileges to entity, modifying simulation parameters or state data, adding/updating S\&S rules, and so on. 

To better understand about the current state of a data object such as \textit{why}, \textit{where}, and \textit{how}, we need to be aware of the complete lineage of process chain (i.e., set of actions performed on data)~\cite{ZAFAR201750}. In this context, provenance-enabled blockchain-based digital twins assure the traceability and integrity of the data, thereby resulting in more informed decisions made by the underlying systems~\cite{suhail2021blockchainbased}. Supporting the digital twin engineering phase through provenance-aware blockchain-based solutions allows monitoring of the transitions in the process knowledge (both specification-based and machine learning-based) through time, outliers, and changes. Moreover, introducing an access control model such as Role-Based Access Control (RBAC) that integrates S\&S rules (as proposed in~\cite{suhail2022towards}) at the digital twin engineering phase lowers security and incident response costs, thereby making subsequent lifecycle phases (such as operation and decommissioning (archived digital twins)) less prone to errors. It is worth noting that it is important to dismantle digital twins or the digital thread while complying with proper media sanitization guidelines. Similarly, recording provenance can help resolve issues during the replication mode cyclic state updates as we have discussed in section~\ref{drawbacks}. However, the challenges associated with the fidelity of digital twins can be partially addressed through access control mechanisms but cannot be handled entirely through a provenance-based solution and is an area of future research.

\subsubsection{Securing lifecycle data}
The lifecycle of a digital twin which spans different phases and includes  multiple stakeholders who perform various tasks on it. Thus, multiparty use of a digital twin affects confidentiality, integrity, availability, and access control \cite{dietz2019distributed}. By establishing a distributed infrastructure, blockchain solves the critical problem of data dissemination across multiple participating entities. Blockchain can manage enterprise policies and rules subject to access rights, i.e., only authorized entities can access, read, and write to the digital twin. Defining access controls may mitigate attacks discussed in section~\ref{rec} and section~\ref{morphing}. Moreover, using its cryptographic strength, blockchain maintains an irrevocable history of digital twin access transactions. It can therefore circumvent the problem of unauthorized data modifications that may invoke other data security-related problems.  

\subsubsection{Role of smart contracts}
Smart contracts allow the execution of code inside a blockchain to automate application-dependent scenarios. When deployed on digital twins, smart contracts can be used to store authorization information for all participating entities~\cite{dietz2019distributed}, track data sharing mechanism~\cite{huang2020blockchain}, represent twin-creation transactions~\cite{shen2021secure}
as surveyed in~\cite{suhail2021blockchainbased}. Additionally, smart contracts, aligned with the predefined conditions, are preferable for scenarios that require automation caused by a change of state, for instance, triggering S\&S rules, invoking PLC functions or due to changing conditions of physical processes or modification of simulation setup parameters. Auditing digital twins by actively or retroactively monitoring smart contract transactions~\cite{suhail2021blockchainbased} can further strengthen the rationale for using smart contracts in digital twins. In this context, auditing helps identify the cause of changes (normal or abnormal). For instance, allowing automation of some of the transaction logic (such as triggering S\&S rules) through smart contracts can help to invoke the appropriate defense mechanisms during the engineering and operation phases.

Considering the dynamism and complexity of ICSs, the impact of autonomous smart contracts might exacerbate before humans understand the situation, validate conditions, and control events. For instance, the system cannot adapt to changing conditions if the rules coded in smart contracts are not dynamically updated. Even dynamic modifications (such as those enabled through AI) may result in the risk of AI making unethical decisions.

\subsubsection{Why may blockchain fail?}
Blockchain mechanisms do not guarantee the trustworthiness of data at the source of the information~\cite{suhail2021securing}. Thus, any weak link in the process chain, either from a physical or virtual environment, can let the attacker enter and carry out malicious activities in the system. For instance, consider a virtual environment that represents IT/OT components of an industrial robotic arm. To command and control the robotic arm (such as turning on/off, speed, controlling joint movements) the PLC and the Human–Machine Interface (HMI) are used whereas the sensors collect the event logs. After gaining access to the virtual environment, an adversary can add or update bad safety and security practices that are then recorded in the blockchain to be used by subsequent processes. Such events may result in Garbage In Garbage Out problem. Furthermore, even if the blockchain is not involved, the cyclic state updates to and from the digital twin environment to the physical process could be enough to cause the damage. To address the GIGO problem, one potential solution is to ensure the trustworthiness of the sources generating the data. To this end, engineering knowledge describing the design specifications at the network/logic layer of the underlying system can be utilized. Generating the network setup of the virtual environment based on technical, topological, and control artefacts can help model the correct behaviour of the physical counterparts~\cite{eckhart2018towards}. Furthermore, engineering knowledge can also serve as a basis for implicit security rules such as defining a safe state based on device benign behavior, cross-validating device data against threshold values, detecting unknown devices or unidentified connections~\cite{eckhart2018towards, suhail2022towards}. 

In an industrial system, a massive amount of real-time data is disseminated to and from digital twins in a continuum to create a digital factory. The reluctance to adopt blockchain in such a digital factory in practice remains an open question. 
To address this concern in blockchain-based digital twins requires further investigation to resolve challenging issues related to scalability, time-varying network delay, time-sensitive tasks (such as real-time remote manipulation), quantum resistance, energy consumption, and integration with legacy systems~\cite{ suhail2021blockchainbased, yaqoob2020blockchain}.

\begin{figure*}[!t]
\centering
\includegraphics[width=5.50in]{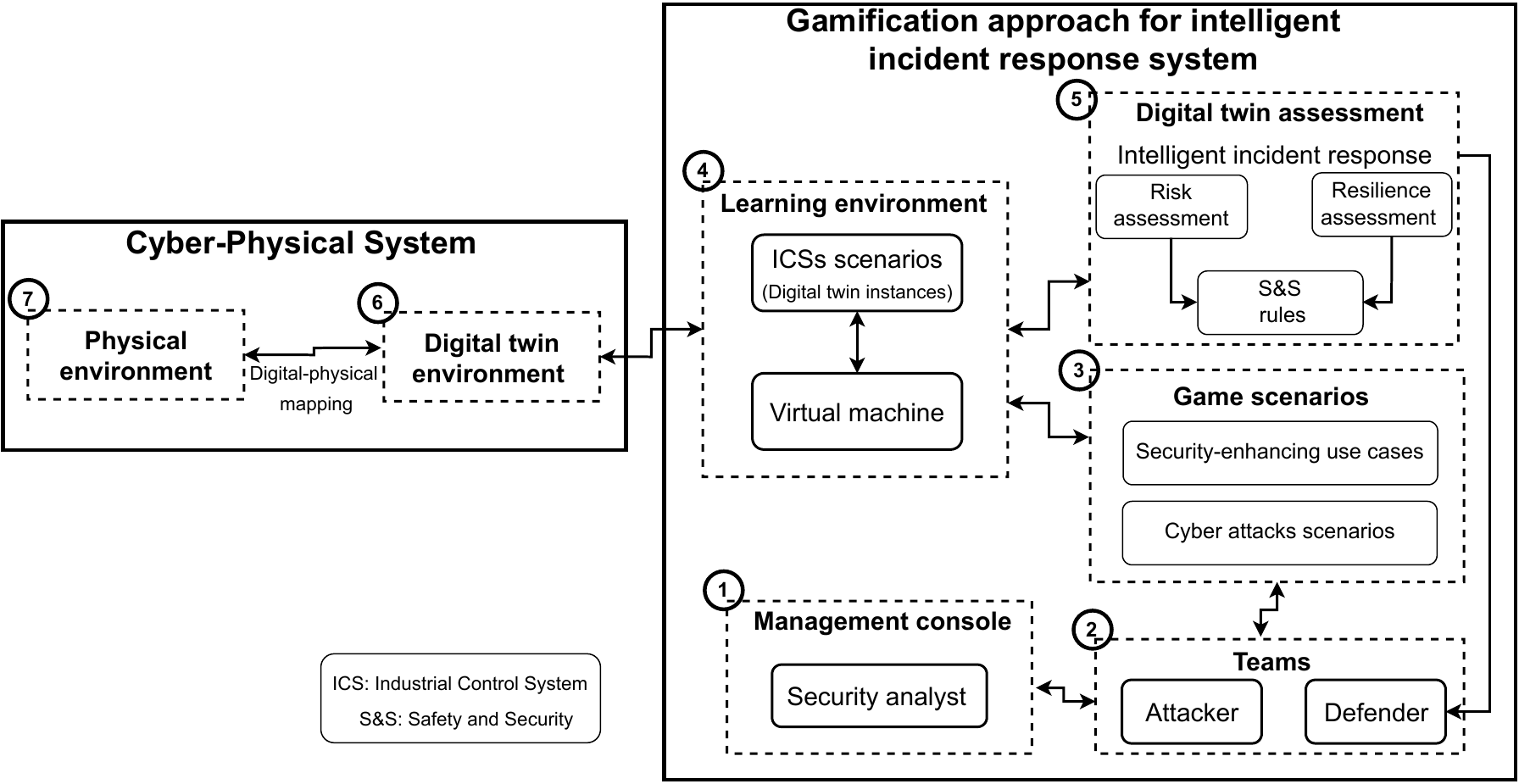}
\caption{Gamification approach for digital twins.}
\label{fig:game}
\end{figure*}

\subsection{Bringing gamification to digital twins security}
Although digital twins operate virtually in an environment distinct from the live system, are prone to attacks. To thwart attacks on digital twins, one potential solution is to assess the security level of digital twins by launching attacks on them. However, such assessment must be performed in an isolated environment without negatively affecting the operation of digital twin modes (especially replication). To this end, we propose a gamification approach that provides twin assessment and a learning environment for security analysts. The following section discusses how the gamification approach can help evaluate the security of digital twins against the attacks specifically discussed in section~\ref{sim-rep}, section~\ref{physical-virtual}, and section~\ref{lateral}.

Gamification is the process of incorporating game mechanics into non-game environments. In cybersecurity, the gamification approach aims to provide security analysts with a controlled, supportive virtual training environment. 
To investigate the resilience of physical processes against attack, determining the potential loss such as disruption of services or machinery breakdown can be gamified.
For instance, the red-blue team cybersecurity exercises~\cite{suhail2021securing}, penetration testing~\cite{Eckhart2019}, Capture-The-Flag (CTF) challenges~\cite{dietz2020unleashing}, or using cyber range to provide hands-on cyber skills and security posture testing~\cite{vielberth2021digital} are among well-known approaches in the existing literature. By simulating attack and defense scenarios, without risking critical infrastructures, such solutions reap the following benefits: (i) we can train security analysts by defining context, environment, and learning objectives to gain practical knowledge and skills during an exercise or challenge, and (ii) we can evaluate the security of digital twins and eventually the physical asset. Leveraging gamification for security-awareness training, can further complement automated security testing of digital twins through incident response which may benefit in lateral movement. Furthermore, it can help to identify attacks during the hybrid approach (discussed in section~\ref{hybrid}) due to the connection between the learning environment and CPS (as Fig~\ref{fig:game} shows). Note that a similar framework can be found in Dietz et al.~\cite{vielberth2021digital}. However, their work focuses more on providing a training environment for security analysts. 

Based on~\cite{vielberth2021digital}, Fig.~\ref{fig:game} showcases the high-level architecture which includes main components of the proposed digital twin-based gamification approach, where our goal is to get the potential defensive solutions by analyzing the attack patterns in a simulated environment (also known as learning environment). Our proposed system comprises two main parts: 1) CPS and 2) gamification approach for intelligent incident response system. The CPS consists of (i) {\it physical environment} (such as a robotic arm) and (ii) {\it digital twin environment} (the virtual copy of a robotic arm). The gamification approach for intelligent incident response system consists of (i) {\it a management console} to assign roles and resources to security analysts, (ii) {\it teams} which include attacker and defender for respective scenarios, (iii) {\it game scenarios} with various attack scenarios, (iv) {\it learning environment} to replicate the digital twin environment for simulation purposes, and (v) {\it digital twins assessment} which analyzes the risks and resilience to automatically get the rules to improve the defensive mechanisms. Next, we describe the use case for the gamification approach. Initially, a security analyst ({\it step~1}) chooses the configuration of the teams ({\it step~2}) made up of attackers or defenders from the management console. 
Next, a security analyst chooses the game scenarios ({\it step~3}) that comprises security-enhancing use cases (for example, security testing, system testing and training) and cyber attacks scenarios (for example, Man-in-the-Middle (MitM), intrusion, altering device configurations or simulation parameters). Then, the learning environment~({\it step~4}) implements the ICS scenarios of different digital twin instances through the virtual machine. The purpose of using a virtual machine is to emulate the functionality of any of the desired digital twin instance without affecting ongoing processes (digital-physical mapping) in CPS. 
In a traditional approach, the learning material for the analyst is provided through videos or instructional texts. The simulated scenario produces log data documenting the operations. The digital twin assessment module~({\it step~5}) is then used to analyze the log data. Based on the incident response playbook, scenario-based learning can be reproduced in terms of the level of difficulty to guide the analyst through several training units (to and from step~4 and step~5). 
 
Determining a viable response to a security incident is vital. Incidents can be a precursor to a future attack or it has already occurred or is currently underway. Investigating incidents provides an opportunity to learn and better prepare for similar incidents in the future. More specifically, it can expose activities (such as an attacker impersonating a legitimate user) associated with lateral movement. 
In this context, the incident response describes the action to be taken based on the type of incident. Responding to a security incident is indispensable to effectively minimize the damage and recovery time while finding and fixing the cause to prevent future attacks. To do so, we introduce an intelligent incident response module that can automate the incident response process. The agents deploy machine/deep learning algorithms (such as Generative Adversarial Networks (GANs)) to execute tasks such as attacking and defending the system. Depending on the game scenario, the attacker launches various types of attacks on the system. In contrast, the defender must detect anomalies and determine the risks associated with the occurrence of anomalies. Both the attacker and defender agents can choose the best attack/response strategy to attack/protect the digital twin environment based on the trained data. For instance, an attacker can determine the probability of a successful attack. To improve the agents’ capabilities, we can train our AI agents in the learning environment based on data used in actual attacks and simulated data. Moreover, an intelligent incident response can construct or update S\&S rules depending on the log data input and resilience assessment sub-modules.
As a result, we can develop the intelligent incident response for the CPS, where the intelligent agent (i.e., detect the anomaly and reconfigure the parameters to mitigate an  attack or to reduce the damage) comes from the gamification process of the learning environment. Based on the intelligent incident response approach, the learning environment result can be directly replicated to the digital twin environment~({\it step~6}) and ultimately the physical environment~({\it step~7}).

To sum up, the proposed gamification approach is best suited for evaluating the security level of both digital modes, i.e., replication and simulation. Furthermore, the required ICS scenario can be virtualized to better understand specific types of attack.  
Since digital twins are the key source of input data, performing security assessments of digital twins can eventually secure the physical system. 

\section{Conclusion}
In his work, we have focused on two aspects: (i) various types of attacks on digital twins, and (ii) defensive strategies to thwart such attacks on digital twins. As attackers are constantly improving their attack techniques, we could adopt the following techniques to limit the damage in the event of a compromise on digital twins.
\begin{itemize}
\item We need intelligence-driven solutions (data analytics and threat intelligence) to collect information on attackers’ behaviors. Threat hunting can use this intelligence to identify forensic artefacts, such as Indicators Of Compromise (IOCs). Recent reports from incident response, for instance, sharing IoCs, i.e., knowledge of suspicious activities and artifacts with the community as suggested by~\cite{skopik2021seven} might also be helpful. During the recovery process, switching off the affected device or service can limit the damage caused by cyberattacks.
\item We should implement provenance-aware blockchain-based solutions to audit digital twins, i.e., track and trace the accountable entity which changed the simulation setup parameters or state data. Provenance data can help reconstruct the process chain to detect and localize the faulty node in the system.
\item We need to develop  a fault-tolerant system. Instead of disconnecting the entire system, we should enable a graceful degradation during which the system enters a fail-safe state 
and maintains a sufficient control of the physical process in case of abnormal incidents. Depending on the priority or severity of the incident, contingency planning is required to identify the root cause of operational disruption or the fraudulent middleman. With short-term remedies and small-scale fixes, we must minimize the probability of incidents and their recovery time.
\end{itemize}
 


\end{document}